\documentclass[preprintnumbers]{revtex4}

\usepackage{graphicx}
\def\mpl{\ifmmode \overline M_{Pl}\else $\overline M_{Pl}$\fi}
\def\ra{\rightarrow}
\def\pairof#1{#1^+ #1^-}
\def\ee{\pairof{e}}
\def\ww{W^+W^-}

\def\dkg{\Delta\kappa_\gamma}
\def\ra{\rightarrow}
\def\wwl{W_{\rm L}W_{\rm L}}

\begin{document}
\bibliographystyle{revtex}

\preprint{SLAC-PUB-9067}

\title{Strong Symmetry Breaking at $\ee$ Linear Colliders}

\author{Timothy L.  Barklow}

\email[]{timb@slac.stanford.edu}
\thanks{Work supported by
Department of Energy contract  DE--AC03--76SF00515.}

\affiliation{Stanford Linear Accelerator Center, 
Stanford University, Stanford, California 94309 USA}

\date{December 18, 2001}

\begin{abstract}
The study of strong symmetry breaking at 
an $e^+e^-$ linear collider with $\sqrt{s}=0.5-1.5$ TeV
is reviewed.    It is shown that processes such as 
$\ee\ra\nu \bar{\nu}\ww$,  
$\ee\ra\nu \bar{\nu} t\bar{t}$,
and  $\ee\ra\ww$ can be used to measure chiral Lagrangian and strong resonance parameters.
The linear collider results are 
compared with those expected from the LHC.
\end{abstract}

\maketitle

\section{Introduction}

Until a Higgs boson with large couplings to gauge boson pairs
is discovered, the
possibility 
of strong electroweak symmetry breaking
must be entertained.  
Without such a particle
the scattering of gauge bosons will become strong at a
scale of order 1~TeV.  
The most commonly studied class of theories 
which deals with this scenario 
is technicolor~\cite{technicolor}. A generic prediction of technicolor theories
is that there is a vector resonance with 
mass below about 2 TeV which unitarizes the $WW$ scattering cross section.
Scalar and tensor resonances are also possible, along with
light pseudo-Goldstone bosons which can 
can be produced in pairs or in association with other particles \cite{Casalbuoni:1998fs}.

Independent of the model, 
the strong interactions of gauge bosons below the threshold for resonance production 
can
be described by an effective chiral Lagrangian
in analogy with $\pi\pi$ scattering below the $\rho$ resonance~\cite{Bagger:1993}~:
\[
{\cal L}_{SB}= {\cal L}^{(2)}+{\cal L}^{(4)}+\cdot\cdot\cdot
\nonumber
\]                                                 
\begin{eqnarray}
 {\cal L}^{(2)} &=& \ \ {v^2\over 4}\, {\rm Tr} D^\mu \Sigma^\dagger
D_\mu \Sigma
   + {{g'}^2 v^2\over 16 \pi^2}\,  b_1\,  ({\rm Tr}
T\, \Sigma^\dagger D_\mu \Sigma )^2 \nonumber \\
&& \qquad \mbox{}+ {gg'\over 16 \pi^2}\,  a_1\, 
{\rm Tr}
 \Sigma B^{\mu\nu}\Sigma^\dagger W_{\mu\nu}\ \ \nonumber \\
&& \nonumber \\
 {\cal L}^{(4)}
 &=& \frac{ \alpha_4 }{16\pi^2}{\rm Tr}
 \Bigl(D_\mu\Sigma^\dagger D_\nu\Sigma\Bigr){\rm Tr}
 \Bigl(D^\mu\Sigma^\dagger D^\nu\Sigma\Bigr)
 +\frac{ \alpha_5 }{16\pi^2}\Bigl[{\rm Tr}
 \Bigl(D^\mu\Sigma^\dagger D_\mu\Sigma\Bigr)\Bigr]^2 \nonumber \\
 & & -ig\frac{ L_{9L\ } }{16\pi^2}{\rm Tr}
 \Bigl(W^{\mu\nu}D_\mu\Sigma D_\mu\Sigma^\dagger\Bigr)
     -ig'\frac{ L_{9R\ } }{ 16\pi^2}{\rm Tr}
 \Bigl(B^{\mu\nu}D_\mu\Sigma^\dagger D_\nu\Sigma\Bigr) \nonumber \ .
\end{eqnarray}
Here $W^{\mu\nu}$ and $B^{\mu\nu}$ are related to the $SU(2)\times U(1)$
gauge fields as in~\cite{Bagger:1993}~, $D_\mu$ is the covariant derivative, 
$g$ and $g'$ are the  $SU(2)\times U(1)$ coupling constants, and $\Sigma$ is composed of the Goldstone boson fields
$w^k$:
\[
 \Sigma = \exp{\Bigl({iw^k\tau^k\over v}\Bigr)} \ ,
\]
where the $\tau^k$ are Pauli matrices and $v=246$~GeV is the Standard Model Higgs vacuum expectation value parameter.
The chiral Lagrangian parameters $a_1$ and $b_1$ are tightly constrainted by precision electroweak data~\cite{Bagger:1999te}.
The terms with coefficients $\alpha_4$ and $\alpha_5$ induce anomalous quartic gauge boson couplings which can be measured 
by observing gauge boson scattering in processes 
such as   $\ee\ra\nu \bar{\nu}\ww$ and $\nu \bar{\nu}ZZ$.
The terms with coefficients  $L_{9L\ }$ and $ L_{9R\ }$ induce
anomalous triple gauge couplings (TGC's) which can be measured
in the reaction $\ee\ra\ww$.


In this paper we summarize strong symmetry breaking signals and the measurement of chiral Lagrangian and strong resonance parameters
at an $\ee$ linear collider (LC) with a center of mass system (CMS) energy in the
range of 0.5 to 1.5 TeV.  Many of the results are taken from the strong symmetry breaking
sections of Ref.~\cite{Abe:2001wn}, which the reader is invited to consult for further details.

\begin{picture}(200,1)
\put(25,-70){\small \textit{Presented at the APS/DPF/DPB Summer Study on the Future of Particle Physics (Snowmass 2001),}} 
\put(110,-82){\small \textit{30 June - 21 July 2001, Snowmass, Colorado, USA}}
\end{picture}

\section{$\ee\ra\nu \bar{\nu}\ww$,\ \   $\nu \bar{\nu}ZZ,\ \  \nu \bar{\nu} t\bar{t}$}

The first step in studying strong $WW$ scattering is
to separate the scattering of a pair of longitudially polarized $W$'s, denoted by $\wwl$,
from transversely polarized $W$'s and background such as $\ee\ra\ee\ww$ and $e^- \bar{\nu}W^+Z$.
Studies have shown that simple cuts can be used to achieve this separation 
in $\ee\ra\nu \bar{\nu}\ww$, $\nu \bar{\nu}ZZ$
at $\sqrt{s}=1000$~GeV, and that the
signals are comparable to those obtained at the LHC~\cite{Barger:1995cn,Boos:1998gw}. 
Furthermore, by analyzing the gauge boson production
and decay angles it is possible to 
use these reactions to measure the chiral Lagrangian parameters   $\alpha_4$ and $\alpha_5$ with an accuracy 
greater than that which can be achieved at the LHC~\cite{Chierici:2001}.

The reaction $\ee\ra \nu \bar{\nu} t\bar{t}$ provides unique access to 
$\ww\ra t\bar t$ since this process is overwhelmed 
by the background $gg\ra t\bar{t}$ at the LHC.  Techniques similar to those employed
to isolate $\wwl\ra \ww, ZZ$ can be used to measure
the enhancement in  $\wwl\ra t\bar{t}$ production~\cite{Barklow:1997wwtt,RuizMorales:1999kz,Han:2001ic,Larios:2000xj}.   Even in the absence 
of a resonance it will be possible to establish a clear signal.  The ratio $S/\sqrt{B}$ is expected to be 12 for 
a linear collider with $\sqrt{s}=1$~TeV, 1000 fb$^{-1}$ and 80\%/0\% electron/positron beam polarization,
increasing to 22 for the same luminosity and beam polarization at $\sqrt{s}=1.5$~TeV.

\section{$\ee\ra\ww$}

Strong gauge boson interactions induce
anomalous TGC's at tree-level:
\begin{eqnarray}
   \kappa_\gamma &=& 1+\frac{e^2}{32\pi^2s_w^2}\bigl(L_{9L}+L_{9R}\bigr) \nonumber
 \\  \kappa_Z  &=& 1+\frac{e^2}{32\pi^2s_w^2}
   \bigl(L_{9L}-\frac{s_w^2}{c_w^2}L_{9R}\bigr) \nonumber
 \\  g_1^Z  &=& 1+\frac{e^2}{32\pi^2s_w^2}\frac{L_{9L}}{c_w^2} \  \nonumber .
\end{eqnarray}
where $\kappa_\gamma$, $ \kappa_Z$, and $g_1^Z$ are TGC's~\cite{Hpzh:1987},
      $s_w^2=\sin^2\theta_w$, and
      $c_w^2=\cos^2\theta_w$.
Assuming  QCD values for the chiral Lagrangian parameters $L_{9L}$ and  $L_{9R}$, $\kappa_\gamma$  is
shifted by  $\dkg \sim -3\times 10^{-3}$.

\begin{table}[]
\begin{center}
\begin{tabular}{l|cc|cc}
     & \multicolumn{4}{c}{error $\times 10^{-4}$} \\
\hline
     & \multicolumn{2}{c|}{$\sqrt{s}=500$ GeV} &  \multicolumn{2}{c}{$\sqrt{s}=1000$ GeV} \\
 TGC & Re & Im & Re & Im \\
\hline
& & & & \\
$g^\gamma_1$ &  15.5    & 18.9     & 12.8     &  12.5     \\
$\kappa_\gamma$ &  \ 3.5    &  \ 9.8    & \ 1.2     &   \ 4.9    \\
$\lambda_\gamma$ & \ 5.4     &  \ 4.1    &  \ 2.0    &   \ 1.4    \\
$g^Z_1$          &  14.1    & 15.6     & 11.0     &  10.7     \\
$\kappa_Z$        & \ 3.8     &  \ 8.1    &  \ 1.4  &  \ 4.2    \\
$\lambda_Z$        &  \ 4.5     & \ 3.5     &  \ 1.7  &  \ 1.2  \\
\hline
\end{tabular}
\caption{Expected errors for the real and imagninary parts of CP-conserving TGCs assuming $\sqrt{s}=500$~GeV, 
${\cal L}=500$~fb$^{-1}$ and  $\sqrt{s}=1000$~GeV, ${\cal L}=1000$~fb$^{-1}$.  The results are for
one-parameter fits in which all other TGCs are kept fixed at their SM values.}
\label{tab:cp-conserving}
\end{center}
\end{table}

The TGCs can be measured by analyzing the $\ww$ production and decay angles in the process $\ee\ra\ww$.
Table~\ref{tab:cp-conserving} contains the estimates of the TGC
precision that can be obtained at $\sqrt{s}=500$ and 1000~GeV for the
CP-conserving couplings  $g^V_1$, $\kappa_V$, and $\lambda_V$.  These estimates are derived
from one-parameter fits in which all other TGC parameters are kept fixed at their tree-level SM values.  
For comparison the LHC with ${\cal L}=300$~fb$^{-1}$ is expected to measure  $\kappa_\gamma$ and $\kappa_Z$
with an accuracy of 0.006 and 0.01, respectively.
The $4\times 10^{-4}$ precision for the TGCs $\kappa_\gamma$ and $\kappa_Z$ at $\sqrt{s}=500$~GeV 
can be interpreted as a precision of $0.26$ for the chiral Lagrangian parameters
$L_{9L}$ and $L_{9R}$.
Assuming naive dimensional analysis~\cite{Manohar:1984md} 
such a measurement 
would provide a $8\sigma$ ($5\sigma$) signal for $L_{9L}$ and $L_{9R}$
if the strong symmetry breaking energy scale were 3~TeV (4~TeV).

When $WW$ scattering becomes strong
the amplitude for $\ee\ra\wwl$ develops a complex form factor $F_T$
in analogy with the pion form factor in $\ee\ra\pi^+\pi^-$~\cite{peskinomnes,Iddir:1990xn}.  To evaluate
the size of this effect the 
following expression for $F_T$ can be used:
\[
 F_T =
         \exp\bigl[{1\over \pi} \int_0^\infty
          ds'\delta(s',M_\rho,\Gamma_\rho)
          \{ {1\over s'-s-i\epsilon}-{1\over s'}\}
         \bigr]
\]
where
\[
\delta(s,M_\rho,\Gamma_\rho) = {1\over 96\pi} {s\over v^2}
+ {3\pi\over 8} \left[ \
\tanh (
{
s-M_\rho^2
\over
M_\rho\Gamma_\rho
}
)+1\right] \ .
\]
Here $M_\rho,\Gamma_\rho$ are the mass and width respectively of a vector resonance in $\wwl$ scattering.
The term 
\[
\delta(s) = {1\over 96\pi} {s\over v^2}
\]
is the Low Energy Theorem (LET) amplitude for $\wwl$ scattering at energies below a resonance.  
Below the resonance, the real part of $F_T$ is proportional to  $L_{9L}+L_{9R}$ and can therefore be
interpreted as a TGC.   The imaginary part, however, is a distinct new effect.

\begin{figure}[tbh] 
\centerline{\includegraphics[angle=-90,clip=,width=9cm]{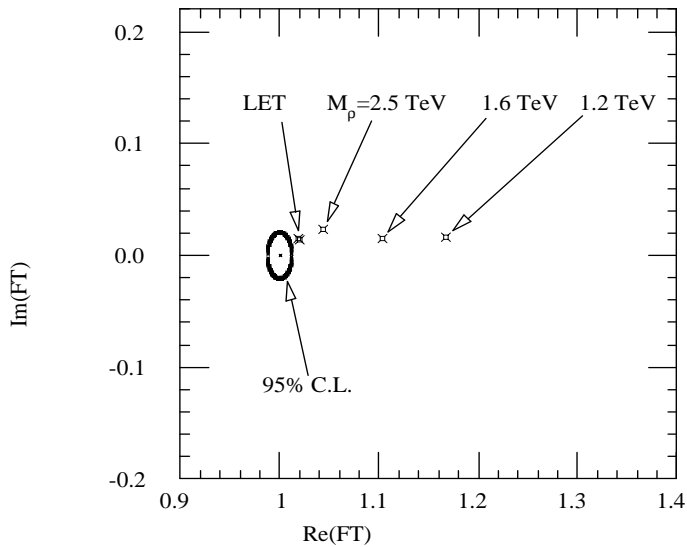}}
\vspace{10pt}
\caption{95\% C.L. contour for $F_T$  for $\sqrt{s}=500$~GeV
and 500~$fb^{-1}$. Values of $F_T$  for various masses $M_\rho$ of a vector resonance in $\wwl$ scattering
are also shown. The $F_T$ point ``LET'' refers to the
case where no vector resonance exists at any mass in strong $\wwl$ scattering.}
\label{fig:fteight}
\end{figure}

\begin{figure}[] 
\centerline{\includegraphics[height=17cm]{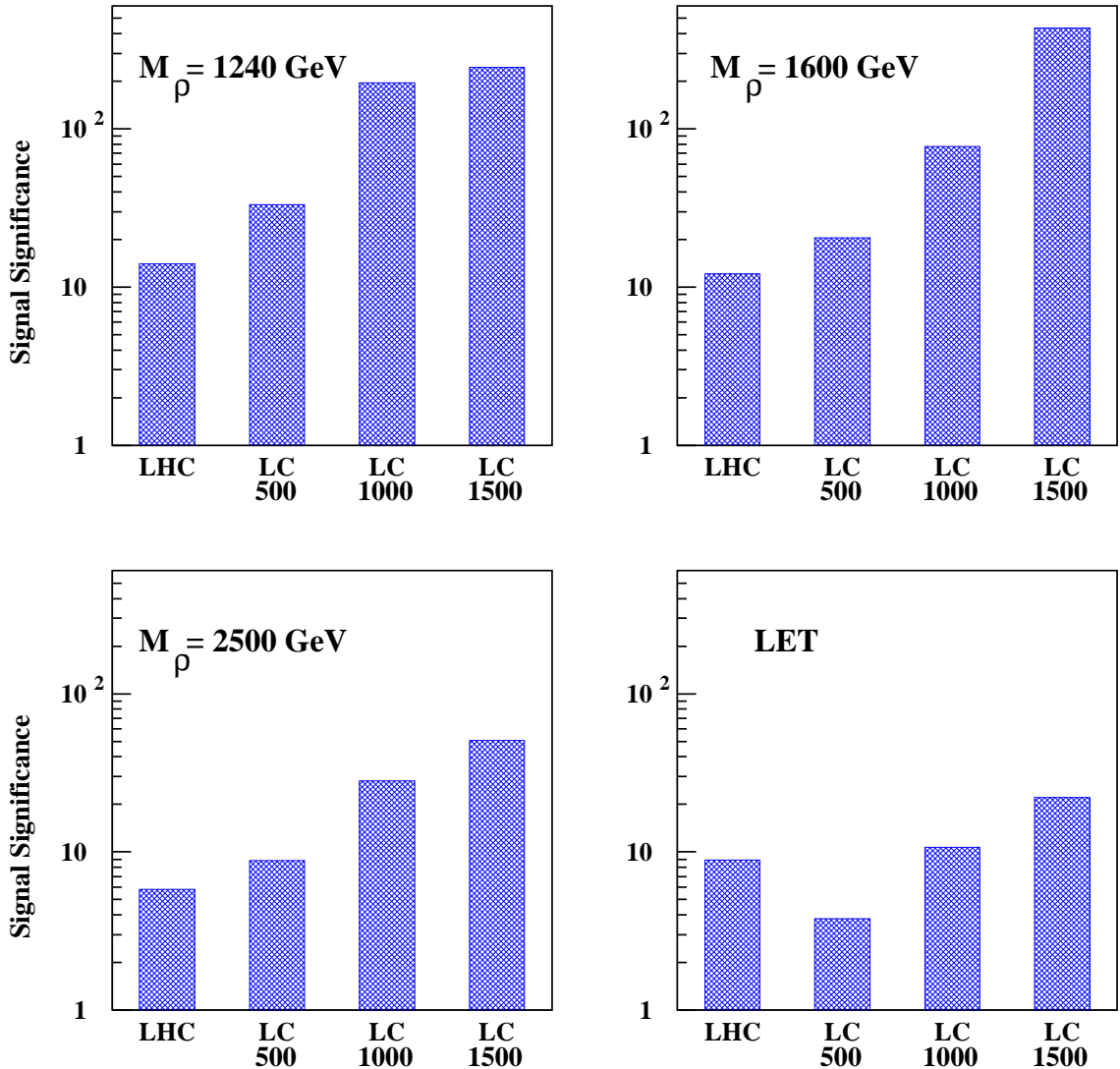}}
\vspace{10pt}
\caption{Direct strong symmetry breaking signal significance in $\sigma$'s for various masses $M_\rho$ of a vector resonance in $\wwl$ scattering.
The numbers below the ``LC'' labels refer to the
center-of-mass energy of the linear collider in GeV.
The luminosity of the LHC is assumed to be 300~$fb^{-1}$, while the lumiosities
of the linear colliders are assumed to be  500, 1000, and 1000~$fb^{-1}$ for 
$\sqrt{s}$=500, 1000, and 1500~GeV respectively.  The lower right hand plot ``LET'' refers to the
case where no vector resonance exists at any mass in strong $\wwl$ scattering.}
\label{fig:strong_lc_lhc}
\end{figure}

The real and imaginary parts of  the form factor  $F_T$ are measured in $\ee\ra\ww$ in the same manner as the TGCs.
The expected 95\% confidence level limits for $F_T$ for $\sqrt{s}=500$~GeV
and a luminosity of 500~$fb^{-1}$ are shown in Figure \ref{fig:fteight}, 
along with the predicted values of $F_T$ for various  masses $M_\rho$ of a vector resonance in $\wwl$ scattering.
The signal significances obtained by combining the results for 
$\ee\ra\nu \bar{\nu}\ww$, $\nu \bar{\nu}ZZ$~\cite{Barger:1995cn,Boos:1998gw} 
with the $F_T$ analysis of $\ee\ra\ww$~\cite{Barklow:2000ci}
are displayed in Fig.~\ref{fig:strong_lc_lhc} along with the
results expected from the LHC~\cite{atlas:1999}.   At all values of the center-of-mass energy
 a linear collider
provides a larger direct strong symmetry breaking signal than the LHC for vector 
resonance masses of 1200, 1600 and 2500~GeV.   
Only when the vector resonance
disappears altogether (the LET case in the lower right-hand plot in Fig.~\ref{fig:strong_lc_lhc} ) does the  direct strong symmetry breaking signal from the $\sqrt{s}=500$~GeV linear collider
drop below the 
LHC signal.   At higher $\ee$ center-of-mass energies the linear collider signal exceeds the LHC signal.

\section{Strong WW Scattering Benchmark Processes}

The Snowmass 2001 working group on experimental approaches at linear colliders used a series of benchmarks to help evaluate
the physics program of a future $\ee$ linear collider \cite{Battaglia:2001}. 
Strong $WW$ scattering in the presence of scalar and vector resonances was simulated using
the model of
Han et al. \cite{Han:2000ic}, 
with resonance masses of 1.0 and 1.5 TeV.  The scalar resonance in this model was basically the SM Higgs.  The 
widths of the vector resonances were 0.055 and 0.077 TeV for resonance masses of 1.0 and 1.5 TeV, respectively.
For non-resonant strong $WW$ scattering the
unitarized K-martrix LET
model\cite{Chanowitz:1993zh} was used.   

When estimating the mass scale reach of the K-matrix LET model and the mass resolution of the resonance model in the presence 
of a scalar (I=0) or  tensor (I=2) resonance,
we use the leading order modifications to the LET cross sections \cite{Barklow:1997nf}
:
\begin{eqnarray}
\sigma(M_0)&=&\left(1+\frac{8}{3}\frac{\hat{s}}{M_0^2}\right)\sigma_{\rm LET} \nonumber \\
\sigma(M_2)&=&\left(1+2\frac{\hat{s}}{M_2^2}\right)\sigma_{\rm LET} \ \ , \nonumber  
\end{eqnarray}
where $M_0$ and $M_2$ are the resonance masses in the $I=0,2$ channels, respectively. 
(The tensor resonance formula is used to estimate LHC mass scale sensitivity.)
For detecting vector resonances we use the technipion form factor, which
to leading order in $s/M_1^2$ is given by
\[
F_T={M_1^2-i\Gamma_1M_1\over M_1^2-s-i\Gamma_1M_1}\ \ \ ,
\]
where $M_1$ and $\Gamma_1$ are the vector resonance mass and width, respectively.
In order to evaluate the vector mass scale reach in the K-Matrix LET model we use the expression
\[
Re(F_T)=1+\Delta_{\rm LET}\ +\frac{s}{M_1^2}\ \ \ ,
\]
where $\Delta_{\rm LET}$ is the contribution to $F_T$ from strong $WW$ scattering in the absence of 
a vector resonance.
The dependence of $\Delta_{\rm LET}$ on the details of the unitarization scheme grows as
$\sqrt{s}$ grows; the systematic uncertainty due to our lack of knowledge of these details
is included in our calculations.

\begin{table}[]
\begin{center}
\begin{tabular}{l|l|c|c||c|c||cc|cc}
    \multicolumn{4}{c||}{} & $M_0=1$ TeV  & $M_0=1.5$ TeV & \multicolumn{2}{|c}{$M_1=1$ TeV}  & \multicolumn{2}{|c}{$M_1=1.5$ TeV} \\
\hline
 Collider & Final & $\sqrt{s}$ & $\mathcal{L}$&
  $\Delta M_0$  &  $\Delta M_0$ & $\Delta M_1$  & $\Delta \Gamma_1$ &  $\Delta M_1$  & $\Delta \Gamma_1$ \\
 & State & TeV & $\rm{fb}^{-1}$ &  GeV  &  GeV  &  GeV  &  GeV   &  GeV  &  GeV  \\ 
\hline
& & & & & & & & & \\
LC & $\ww$ & 0.5 & 500 & --  & --  &    5.8      &  19.0   &   27.6     &  90  \\
LC & $\ww$ & 1.0 & 1000  &      89  &      249        &   0.01     &   0.03  &   4.0     &  13.5  \\
LC & $\ww$ & 1.5 & 1000  &      14  &      46   &     --      &    --    &    0.04     &  0.15  \\
& & & &  &  & & & & \\
\hline
\end{tabular}
\caption{Expected error $\Delta M_0$ for the mass of
a scalar resonance, and expected errors $\Delta M_1$ and  $\Delta \Gamma_1$ for the mass and width, respectively, of
a vector resonance.  Results are shown for vector resonances of mass 1.0 and 1.5 TeV.}
\label{tab:vector-res}
\end{center}
\end{table}

The expected errors for the mass of the scalar resonances are shown in Table~\ref{tab:vector-res}, along
with the expected errors for the mass and width of the 
vector resonances.  The measurement of the scalar mass $M_0$ is assumed to come solely from the measurement of the cross section $\sigma$,
with $\sigma(M_0)$ defined above.  For the measurement of the scalar mass there is a clear advantage in going to the 
higher CMS energy of 1.5~TeV.
In contrast, 
the masses and widths of the vector resonances are measured very well at all CMS energies.  Even the most poorly measured
vector resonance parameter -- the width of the 1.5 TeV resonance at $\sqrt{s}=0.5$~TeV -- is measured with an 
accuracy of 6\% .
At $\sqrt{s}=1.0$ and 1.5 TeV the vector mass and width resolutions are typical of
an  $\ee$ collider sitting on top of the resonance.

Results for the K-matrix LET model are shown in Table~\ref{tab:kmatrix-res}.   The signal significance
is displayed along with the 95\% C.L. mass scale limits in the  $I=0,1$ isospin channels.
For comparison, results are also shown for the LHC in the $I=2$ channel \cite{atlas:1999}.  
The tensor mass scale lower limit from the LHC is comparable to the scalar mass scale limits from the
LC.   Not suprisingly, the largest mass scale limits are the
vector limits obtained in $\ee\ra\ww$.  Note that the
vector mass scale lower limit $M_1$ does not improve as the CMS energy is raised from 1.0 to 1.5~TeV: this
is due to the systematic uncertainity in the calculation of  $\Delta_{\rm LET}$,
which becomes important near $\sqrt{s}=1.5$~TeV.  The only way to reduce this particular systematic uncertainty is to actually
do strong scattering experiments at the LHC and at an $\ee$ LC.

\begin{table}[]
\begin{center}
\begin{tabular}{l|l|c|c||c|c|c|c}
 Collider & Final & $\sqrt{s}$ & $\mathcal{L}$&
Signal & $M_0$ (TeV) &  $M_1$ (TeV) &  $M_2$ (TeV) \\
 & State & TeV & $\rm{fb}^{-1}$ &  signif.   &  95\% C.L.  &  95\% C.L.  &  95\% C.L.    \\ 
\hline
& & & & &  & &  \\
LC & $\ww$ & 0.5 & 500 &  $3\sigma$  &  --  &   4.8   & --  \\
LC & $\ww$ & 1.0 & 1000  & $7\sigma$  &  --  &   6.4  & --   \\
LC & $\ww$ & 1.5 & 1000 & $8\sigma$ & --   &   6.4   & --   \\
& & & & &  & &  \\
LC & $\nu\bar{\nu}\ww,ZZ$ & 1.0 & 1000 &    $7\sigma$      &    1.7            &    --      &   --       \\
LC & $\nu\bar{\nu}\ww,ZZ$ & 1.5 & 1000 &  $20\sigma$       &   4.3           &   --        &     --     \\
& & & & &  &   &  \\
LHC & $qq W^+W^+$ & 14 & 300 &  $9\sigma$  &    --        &    --        & 3.0 \\
& & & & &  & &   \\
\hline
\end{tabular}
\caption{Signal significance and 95\% C.L. mass scale lower limits 
for the LET model with the K-matrix unitarization scheme.  The mass scales $M_0,M_1,M_2$
correspond to structure in $WW$ scattering in the I=0,1, and 2 isospin channels, respectively.}
\label{tab:kmatrix-res}
\end{center}
\end{table}


\section{Summary}

Studies of strong electroweak symmetry breaking are enhanced by 
an $\ee$ linear collider with $\sqrt{s}=0.5-1.5$~TeV. 
An LC complements a hadron collider nicely in providing 
better measurements of the chiral Lagragian parameters  $L_{9L\ }$ and $ L_{9R\ }$
which affect triple gauge boson vertices.
Also, the LC provides competitive measurements of
the  chiral Lagragian parmeters $\alpha_4$ and $\alpha_5$ which affect quartic gauge boson vertices.

A non-resonant strong symmetry breaking signal will be slightly larger  at a $\sqrt{s}=1.0$~TeV LC than at 
the LHC, and will be significantly larger if the $\ee$ CMS energy is raised to $\sqrt{s}=1.5$~TeV.
Less energy is required for strong vector resonance detection.
A $\sqrt{s}=0.5$~TeV LC provides a larger vector resonance signal than the LHC
for masses up to at least 2.5~TeV.
The mass and width of a strong vector resonance
can be measured at a LC with at least a  few percent accuracy, even when the resonance lies well above the $\ee$  CMS energy.

Another important aspect of strong symmetry breaking is the study of $\ww\ra t \bar t$.  This
reaction can probably only be studied at a LC.  Good strong symmetry breaking signals can be obtained in this channel
at a LC, and these results should prove valuable in understanding electroweak symmetry breaking in the fermion sector.

Finally, we note that the systematic errors in signal and background 
calculations will be smaller at a LC than at a hadron collider, since the production mechanisms and
backgrounds are limited to electroweak processes.  However, we cannot at this time
quantify this advantage since  detailed studies of theoretical systematic errors in strong $WW$ scattering
have not been performed for either the LHC or the LC.  This issue  could be important given 
the size of some of the strong symmetry breaking signals and the paucity of sharp resonances in 
many strong symmetry breaking scenarios.

%
\def\MPL #1 #2 #3 {Mod. Phys. Lett. {\bf#1},\ #2 (#3)}
\def\NPB #1 #2 #3 {Nucl. Phys. {\bf#1},\ #2 (#3)}
\def\PLB #1 #2 #3 {Phys. Lett. {\bf#1},\ #2 (#3)}
\def\PR #1 #2 #3 {Phys. Rep. {\bf#1},\ #2 (#3)}
\def\PRD #1 #2 #3 {Phys. Rev. {\bf#1},\ #2 (#3)}
\def\PRL #1 #2 #3 {Phys. Rev. Lett. {\bf#1},\ #2 (#3)}
\def\RMP #1 #2 #3 {Rev. Mod. Phys. {\bf#1},\ #2 (#3)}
\def\NIM #1 #2 #3 {Nuc. Inst. Meth. {\bf#1},\ #2 (#3)}
\def\ZPC #1 #2 #3 {Z. Phys. {\bf#1},\ #2 (#3)}
\def\EJPC #1 #2 #3 {E. Phys. J. {\bf#1},\ #2 (#3)}
\def\IJMP #1 #2 #3 {Int. J. Mod. Phys. {\bf#1},\ #2 (#3)}
\def\JHEP #1 #2 #3 {J. High En. Phys. {\bf#1},\ #2 (#3)}

\end{document}